\documentclass[10pt]{amsart}
\usepackage{epsf}
\usepackage{amsmath}
\usepackage{amssymb}
\usepackage{cite}

\begin{document}
\def\vac{{\text{\O}}}

\parskip 3mm

\title{Cyclically coupled spreading and pair annihilation}
\author{Haye Hinrichsen}
\address{Theoretische Physik, Fachbereich 10,
     Gerhard-Mercator-Universit\"{a}t Duisburg,
     47048 Duisburg, Germany}
\date{April 19, 2000}

\begin{abstract}
Recently it has been shown that the transition of the
1+1-dimensional annihilation-fission process
$2X$$\rightarrow$$3X$, $2X$$\rightarrow$$\vac$ exhibits an unusual
type of nonequilibrium critical behavior. The phenomenological
properties of critical clusters are characterized by two different
dynamic modes for spreading and diffusion. In order to describe
the interplay of these modes, we introduce an effective model which involves
two species of particles $A$ and $B$. The $A$-particles perform an
ordinary directed percolation process while the $B$ particles
diffuse and annihilate. Both subsystems are cyclically coupled by
particle transmutation $A \leftrightarrow B$. The resulting
critical behavior is in many respects similar to the one observed
in the annihilation-fission process.
\end{abstract}
\maketitle

\section{Introduction}
%
The study of phase transitions into absorbing states is a
fascinating field of nonequilibrium statistical
physics~\cite{MarroDickman99}. 
Such phase transitions can
be observed in models for the spreading of some (generally
non-conserved) agent as a result of a competition between local
reproduction and decay. If the rate for reproduction is
sufficiently high, the system is able to maintain a fluctuating
active phase where the stationary concentration of the spreading
agent is positive. On the other hand, if the decay process
dominates, the concentration of the spreading agent decreases and
tends to zero. Eventually the system becomes trapped in an
absorbing state from where it cannot escape. An interesting
situation emerges at the borderline between survival and
extinction. Here the system undergoes a nonequilibrium phase
transition which is characterized by non-trivial critical
behavior. It is believed that transitions into absorbing states
can be categorized into a finite number of universality classes.
Typically each class is associated with certain symmetry
properties of the dynamics.

At present two universality classes are firmly established. The
first and most prominent one is the universality class of {\em
directed percolation}
(DP)~\cite{BroadbentHammersly57,Kinzel83,Hinrichsen00}. 
The DP class is characterized by the 
{\em absence} of symmetries (apart from
conventional symmetries such as translation and reflection
invariance) and covers a wide range of
models~\cite{Janssen81,Grassberger82}. Roughly speaking, DP models
follow the reaction-diffusion scheme $X \rightarrow 2X$, $X
\rightarrow \vac$. In addition, there has to be a nonlinear mechanism
which limits the particle density. In `fermionic' models with  at most 
one particle per site, the density is limited automatically. In
`bosonic' models, allowing for infinitely many particles per site,
this mechanism has to be implemented by adding the reaction $2X
\rightarrow X$. It is important to note that DP is a {\em unary}
spreading process, i.e., individual particles are able to
reproduce and destruct themselves.

Some time ago Grassberger {\it et~al.}~\cite{GKT84,Grassberger89b}
discovered a second universality class which can be considered as a
generalization of directed percolation from one to two absorbing
states related by an {\em exact} $Z_2$-symmetry (DP2).
It comprises various models, including nonequilibrium Ising
models~\cite{Menyhard94}, certain monomer-dimer
models~\cite{KimPark94}, as well as generalized versions
of the Domany-Kinzel model and the contact process~\cite{Hinrichsen97}. 
In 1+1 dimensions it is possible to
regard kinks between differently oriented absorbing domains as
particles. By definition, these particles evolve according to a
parity-conserving dynamics, exemplified by branching-annihilating
random walks with two offspring $X \rightarrow 3X$, $2X
\rightarrow \vac$ \cite{TakayasuTretyakov92,ALR93,CardyTauber96}.
For this reason, the universality class is often referred to as the
{\em parity-conserving} (PC) class. However, it should be noted that
the PC class and the DP2 class are different in higher dimensions.

Three years ago, Howard and T\"auber~\cite{HowardTauber97} raised
the question whether there might exist a third universality class
of models with a phase transition from `real' to `imaginary' noise.
As a prototype, they introduced the annihilation-fission (AF) 
process $2X \rightarrow 3X$, $2X \rightarrow \vac$ with
single-particle diffusion. Obviously, this process is
neither parity-conserving nor invariant under any other
unconventional symmetry transformation. In contrast to DP and
DP2, the AF model is a {\em binary} process, i.e., only pairs of
particles can decay or reproduce themselves. The critical
properties at the transition are still poorly understood. 
Recently, Carlon {\em et. al.} analyzed the transition using
density matrix renormalization group techniques~\cite{CHS99}.
Estimating the critical exponents they arrived at the conclusion
that the transition of the AF process should belong to the DP2
universality class, although there is no $Z_2$ symmetry
or parity conservation law. However, as pointed out
in Ref.~\cite{Hinrichsen00b}, various physical arguments 
suggest that the AF process might belong to an independent 
universality class which has not been investigated before. 

Since the universality class could not be 
identified so far, it is of interest to understand the most 
salient features of the transition in the AF process
from a descriptive point of view.  
To this end, we introduce an effective model
which separates the dynamics of pairs and solitary
particles in the AF process by introducing two species of
particles $A$ and $B$. The $A$'s perform an ordinary DP process
while the $B$'s are subjected to an annihilating random walk. Both
subsystems are cyclically coupled by particle transmutation. It is
shown that this three-state model exhibits a nonequilibrium phase
transition which is similar to the one observed in the AF process.

The following Section briefly summarizes the phenomenological 
properties of the AF process. In Sec.~\ref{Interpretation} we
demonstrate that the critical behavior is governed by two competing
dynamic modes for spreading and diffusion. Based on this
interpretation we introduce an effective model involving
two species of particles (see Sec.~\ref{Model}). In
Sec.~\ref{Numerics} the critical exponents of this model are estimated
by Monte Carlo simulations. The article ends with several concluding 
remarks in Sec.~\ref{Conclusions}.

%
%
\section{Phenomenological properties of the annihilation-fission process}
\label{Summary}
%
%
Assuming the usual scaling picture for phase transitions into
absorbing states, we expect the AF process
to be characterized by four critical exponents $\beta$, $\beta'$,
$\nu_\perp$, and $\nu_\parallel$. The first one is associated with the
field-theoretic annihilation operator and describes the behavior
of the stationary density $\rho_{\text{stat}}\sim(p-p_c)^\beta$
close to the transition. The exponent $\beta'$ is associated with
the creation operator and plays a role whenever initial conditions are
specified. For example, the survival probability of a cluster
grown from a single seed involves the exponent
$\delta'=\beta'/\nu_\parallel$. The other two exponents are
related to the spatial and temporal correlation lengths
$\xi_\perp\sim|p-p_c|^{-\nu_\perp}$ and
$\xi_\parallel\sim|p-p_c|^{-\nu_\parallel}$, respectively. It
should be noted that in the case of DP the two order parameter
exponents $\beta$ and $\beta'$ coincide because of a duality
symmetry (see e.g. Ref.~\cite{Hinrichsen00}). In
the present case, however, they turn out to be different.

Using a density matrix renormalization group approach, Carlon {\it
et. al.} estimated the critical exponents of the AF process for
various diffusion rates  $0.1 \leq d \leq 0.2$. Since their
estimates $z=1.73\ldots 1.81$  and $\beta/\nu_\perp=0.46 \ldots
0.5$ were in fair agreement with the numerical values of DP2
exponents (see Table 1), 
they concluded that the AF process should belong to the
DP2 universality class, although there is no $Z_2$-symmetry or
parity conservation law in the system. This conclusion, however,
collides with the general believe that critical phenomena are
determined by their symmetry properties or, equivalently, by their
associated field theories. Without the required symmetries on the
microscopic level, a special mechanism would be necessary in order to
restore these symmetries effectively on large scales. As there
is no such mechanism, it is near at hand to expect
that the AF process does not belong to the DP2 class. To support
this point of view, preliminary Monte-Carlo simulations were
presented in Ref.~\cite{Hinrichsen00c} and later improved
by Grassberger and \'Odor~\cite{Private}. Because of
strong corrections to scaling, the estimates for the critical
exponents depend on the numerical effort.
The estimates for $\delta=\nu_\perp/\nu_\parallel$, for example, are
scattered over the range $0.25\ldots 0.29$ and seem to
decrease with increasing simulation time. A similar tendency 
was observed for the exponents $z$ and $\beta$.
Moreover, it is not yet fully clear to what extent the 
exponents depend on the diffusion rate $d$. A tentative list
of critical exponents, including recent results obtained by
simulations on parallel computers~\cite{Odor00}, is given
in Table 1. 

\begin{table}
\begin{tabular}{|c||c|c|c|c|c|}
\hline class/model & $\beta$ & $\delta$ & $z$ & $\delta'$ & $\eta$
\\ %
\hline \hline DP & 0.2765 & 0.1595 & 1.581 & 0.1595 & 0.3137\\ %
\hline DP2/PC & 0.92(2) & 0.286(2) & 1.74(2) & 0.286(2) or 0 & 0 or 0.286(2) \\ %
\hline 
\hline AF d=0.1 & $0.57\ldots0.62$ & $0.25\ldots 0.29$ & 1.67\ldots 1.83 & $0.12\ldots 0.15$ & $\approx 0.10$ \\ %
\hline AF d=0.5 & $0.38\ldots0.42$ & $0.21\ldots 0.23$ & $\approx 1.7$   & $\approx 0.145$   & $\approx 0.23$ \\ %
\hline AF d=0.9 & $0.38\ldots0.40$ & $\approx 0.20   $ &  --	         & --                 & $\approx 0.48$ \\ %
\hline \hline present model & $0.38(6)  $ & $0.21(1)$         & $1.75(10)$      & $0.15(1)$	     & $0.21(2)$ \\ %
\hline
\end{tabular}
\vspace{2mm} \caption{\label{Tab} Estimates for the critical
exponents in 1+1 dimensions.}
\end{table}

In order to understand the transition in the annihilation-fission
process from a phenomenological point of view it is helpful to
analyze the spatio-temporal structure of critical clusters. To
this end we introduce a novel type of scale-invariant
space-time plot which can be used to visualize the scaling
properties of critical clusters in systems with absorbing states.
Starting with a localized seed (a pair of particles) at the origin
and simulating the spreading process up to $10^6$ time steps, the
rescaled position of the particles $x/t^{1/z}$ is plotted against
$\log_{10} t$, where $z=\nu_\parallel/\nu_\perp$ is the dynamic
exponent of the process under consideration (see Fig.~\ref{FIGDEMO}). 
By rescaling the spatial coordinate $x$, 
the cluster is confined to a strip of finite
width. Compared to linear space-time plots the scale-invariant
representation of a cluster has several advantages. On the one
hand, it is possible to survey more than four decades in time. On
the other hand, the plot provides a simple visual check of scaling
invariance. Roughly speaking, scaling invariance is fulfilled if
the cluster's appearance is time-independent, i.e.,
spatio-temporal patterns should look similar in the upper and
lower parts of the figure. It is needless to say that 
this visual check does not replace an 
accurate quantitative analysis. Nevertheless such a 
scale-invariant plot may improve the intuitive understanding 
of the asymptotic long-time behavior and may also help to identify 
relevant and irrelevant  contributions.

Let us first consider a critical cluster of a branching-annihilating
random walk with two offspring (see left part of Fig.~\ref{FIGDEMO}).
Obviously, this process is characterized by an ongoing competition
between particle reproduction and decay. Contrarily,
the annihilation-fission process admits undisturbed random walks of
solitary particles over long distances. As shown in the middle of
Fig.~\ref{FIGDEMO}, this leads to a very different visual
appearance of the cluster. It is important to note that these
differences persist as time proceeds. However, if both processes
were to converge to the same type of long-range critical behavior
as suggested in~\cite{CHS99}, we would expect 
the clusters to become increasingly similar in the
lower part of the figure. As there is no indication of
such a convergence, Fig.~\ref{FIGDEMO} supports the
viewpoint of Refs.~\cite{Hinrichsen00b,Odor00}
that the AF process might represent a new type of 
nonequilibrium critical behavior. The corresponding universality class 
should be characterized by the following properties:

\begin{figure}
\epsfxsize=155mm \vspace{5mm} \centerline{\epsffile{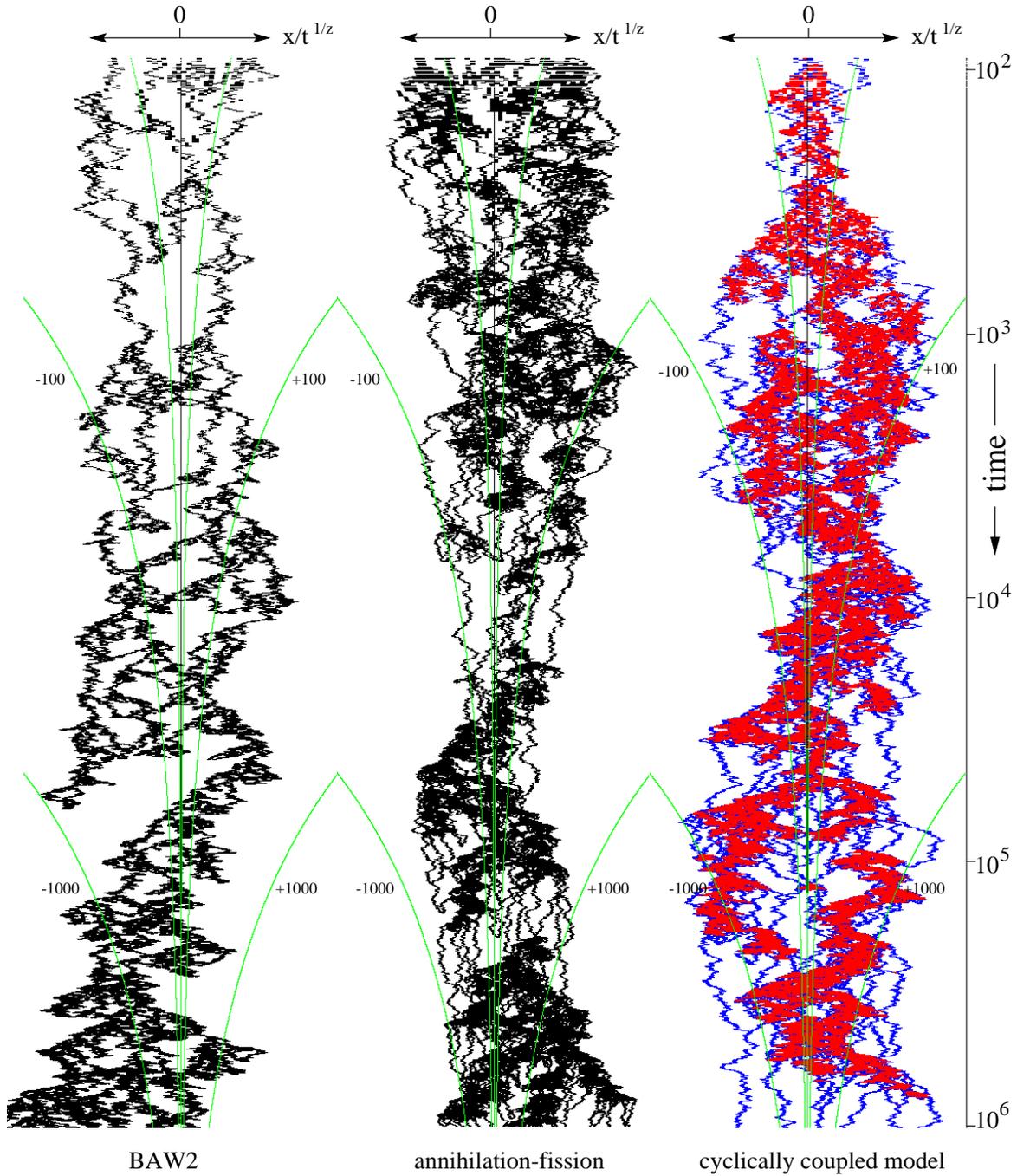}}
\vspace{2mm} \caption{ \label{FIGDEMO} Critical clusters generated
from a single seed at the origin in a scale-invariant
representation. The rescaled position $x/t^{1/z}$ of the particles
is plotted against time on a logarithmic scale. The curved lines
mark the positions $x=\pm 10$, $\pm 100$, and $\pm 1000$, respectively. 
Left: Branching-annihilating random walk with two offspring (BAW2).
Middle: Annihilation-fission process (also called pair contact process
with diffusion). Right: Cyclically coupled spreading and
annihilation processes as an effective model for the AF process. The
figure is explained in the text.}
\end{figure}
\newpage
\begin{enumerate}
\item Single particles diffuse but do not react. %
\item Reproduction requires two particles to meet at neighboring sites.
\item Particles are removed if {\em at least} two particles meet at neighboring sites.
\item There is no unconventional symmetry (such as parity conservation).
\item There is no frozen disorder in the system.
\item There is a mechanism limiting the density of particles.
\end{enumerate}
Consequently, many other processes, such as the
fission-coagulation model $2X\rightarrow 3X$, $2X\rightarrow X$ and
the reaction-diffusion process $2X\rightarrow 3X$, $3X\rightarrow X$
are expected to exhibit the same type of nonequilibrium critical
behavior.

\section{Interpretation as a spreading process with two species of particles}
\label{Interpretation}
%
%
%
%
\begin{figure}
\epsfxsize=110mm \centerline{\epsffile{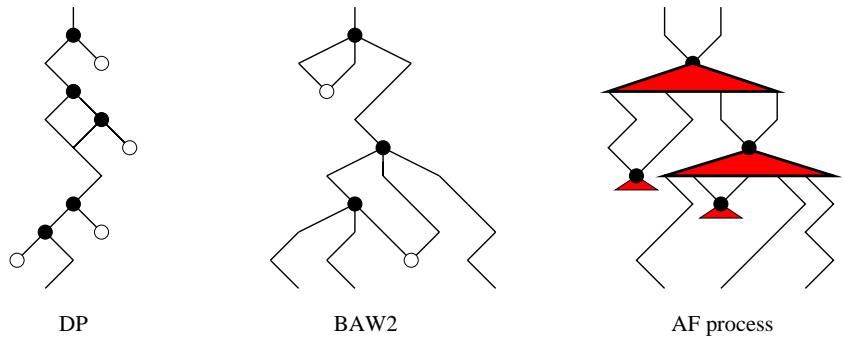}} \vspace{2mm}
\caption{ \label{FIGCARTOON} %
Cartoon of three different 1+1-dimensional spreading processes.
Solid (hollow) circles denote reproduction (decay) of particles.
Left: Directed percolation. Middle: Parity-conserving branching
annihilating random walk with two offspring. Right:
Schematic drawing of the annihilation-fission process
where spreading occurs in avalanches (represented as
triangles) followed by diffusion of solitary
particles (see text).}
\end{figure}
To what extent is the transition in the AF model different from ordinary
DP and DP2 transitions? As Fig.~\ref{FIGDEMO} suggests, there are
{\em two separate dynamic modes} for spreading and diffusion. The
spreading mode is characterized by sudden avalanches with
a high density of particles. Here the dynamic processes are
dominated by interacting {\em pairs} of particles. Once an
avalanche has stopped, the system enters the diffusive mode,
in which solitary particles perform a simple random walk.
When two of them meet at neighboring sites, they may release
a new avalanche, as illustrated in Fig.~\ref{FIGCARTOON}.
The asymptotic critical behavior at the transition will depend on
the relevance of the two dynamic modes. In principle there are
three possibilities. If the spreading mode becomes dominant
we expect a crossover to DP. Conversely, if
the diffusive mode governs the asymptotic regime,
we expect a purely diffusive behavior with the dynamical
critical exponent $z=2$.
However, Fig.~\ref{FIGDEMO} strongly suggests that both modes are
equally important and balance one another as time proceeds. In fact,
in the rescaled representation the typical spatio-temporal patterns
do not change over four decades in time. 

\begin{figure}
\epsfxsize=85mm \centerline{\epsffile{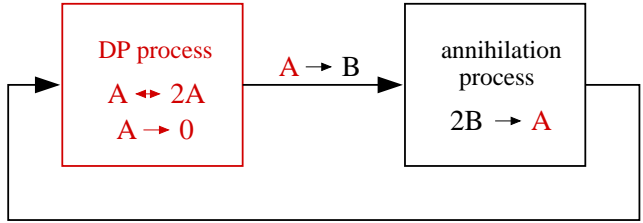}} \vspace{2mm}
\caption{\label{FIGCYCLIC} Cyclically coupled directed percolation
and annihilation.}
\end{figure}

In order to investigate this transition in more detail,
we suggest a phenomenological explanation of the
observed spatio-temporal patterns. The basic idea is to describe
the two dynamic modes in terms of two separate reaction-diffusion
processes involving two different 
species of particles $A$ and $B$.
The spreading mode is governed by the dynamics of $A$-particles.
Roughly speaking, the $A$-particles can be thought of as 
representing {\em pairs} of particles in the original model. 
Obviously, there is no parity conservation law for the number of pairs,
hence the $A$-particles in the new model evolve effectively in
the same way as in an ordinary DP process.
During the avalanche, several $A$'s transmute into $B$'s.
Therefore, once the avalanche has stopped, several $B$-particles
are left behind. These $B$-particles in turn perform a simple
random walk, representing solitary particles in the original
model. When two $B$-particles meet, they may trigger a new avalanche
of $A$-particles. The corresponding reaction-diffusion scheme reads
\begin{equation}
\label{ABSchme}
A \leftrightarrow 2A \,, \quad %
A \rightarrow \vac / B  \,, \quad %
2B \rightarrow A  \,.
\end{equation}
Thus the  model consists of two subsystems {\bf A} and {\bf B}.
Subsystem {\bf A} is a DP process $A
\leftrightarrow 2A, \, A \rightarrow \vac$  which is
coupled via transmutation $A \rightarrow B$ to subsystem {\bf B}.
In this subsystem the $B$-particles
diffuse until they annihilate and release a
new avalanche of $A$-particles. Thus, the reaction-diffusion model
(\ref{ABSchme}) can be interpreted as a cyclically coupled sequence of a
DP and a pair annihilation process, as sketched in
Fig.~\ref{FIGCYCLIC}. It should be pointed out that
Figs.~\ref{FIGCARTOON}-\ref{FIGCYCLIC} are over-simplified in the sense
that they suggest the existence of strictly separated dynamic modes.
In reality, however, the two modes are not completely separated, 
rather they are entangled and sustain 
each other in a intricate manner. 
Nevertheless it is the hope of the present study that
the AF-process and the simplified reaction-diffusion 
scheme~(\ref{ABSchme}) exhibit at least qualitatively a similar
type of nonequilibrium critical behavior.
%
%
%
\section{A lattice model for cyclically coupled directed
percolation and pair annihilation}
\label{Model}
%
%
In order to study the process~(\ref{ABSchme}) quantitatively,
we introduce a three-state model on a square lattice
with random-sequential updates which is defined by the
following dynamic rules:
\begin{center}
\begin{tabular}{llll}
reproduction: & $\vac A \rightarrow AA$ & with rate  & $p/2$ \\ %
             & $ BA  \rightarrow AA$ && $p/2$ \\ %
             & $A\vac  \rightarrow AA$ && $p/2$ \\ %
             & $AB   \rightarrow AA$ && $p/2$ \\[2mm] %

decay: & $A \rightarrow \vac$ && $(1-p)(1-\tau)$ \\[2mm] %
transmutation: & $ A \rightarrow B$ && $(1-p)\tau$ \\[2mm] %
diffusion: & $\vac B \leftrightarrow B\vac $ && $D/2$ \\[2mm] %
annihilation: & $BB \rightarrow A\vac $& & $r/2$ \\ %
 & $BB \rightarrow \vac A$ && $r/2$  %
\end{tabular}
\end{center}
\begin{figure}
\epsfxsize=65mm
\vspace{4mm}
\centerline{\epsffile{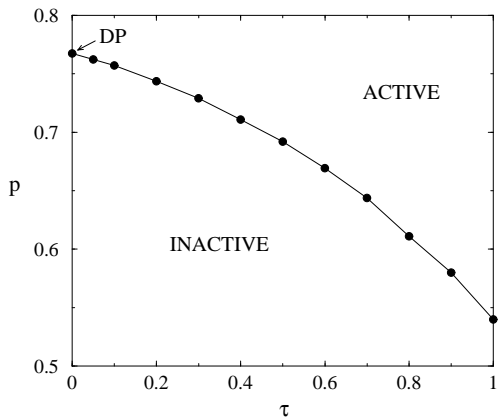}}
\vspace{-4mm}
\caption{ \label{FIGPHASEDIAG} Phase diagram of the lattice
model defined in Sec.~\ref{Model}.}
\end{figure}
In the following the numerical analysis will be restricted to the case
$r=D=1$. Thus, the model is controlled by two parameters,
namely the rate for particle reproduction $p$ and the
transmutation rate $\tau$. Obviously, the model has two absorbing
states, i.e., the empty lattice and the state with a single
diffusing $B$-particle. The phase diagram in
Fig.~\ref{FIGPHASEDIAG} comprises two phases. For low values of
$p$ and $\tau$, the system is in the inactive phase where it
approaches one of the two absorbing states. If $p$ and $\tau$ are
sufficiently large, an active steady state with non-vanishing
particle densities $\rho_A$ and $\rho_B$ exists on the infinite
lattice. Here we are interested in the critical behavior at the
phase transition line. It should be noted that the case $p=0$ is
special. In this case the spreading process does not generate $B$
particles. Hence, starting with $A$-particles, the critical
behavior belongs to the DP universality class. For $p>0$, however, we
observe non-DP critical behavior. In fact, an ordinary DP 
transition seems to be unlikely since the inactive 
phase is charactrized by an algebraic 
decay of the particle density. Similarly, we can rule out the
possibility of a DP2 transition since there is neither a 
$Z_2$-symmetry nor a parity conservation law in the model.

\section{Numerical results}
\label{Numerics}
%
%
%
\begin{figure}
\epsfxsize=120mm \vspace{2mm} \centerline{\epsffile{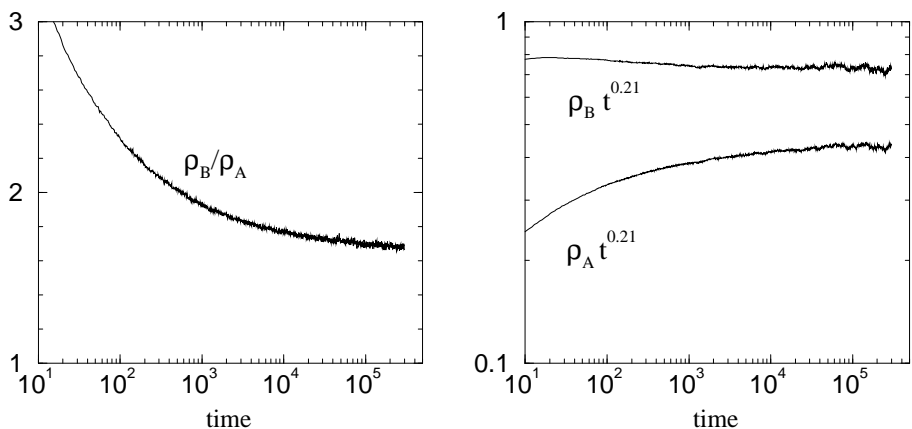}}
\caption{ \label{FIGDECAY} Simulations at the critical point
$\tau=0.5$, $p_c=0.6920$ starting with random
initial conditions. Left: The ratio of the particle densities
$\rho_B/\rho_A$ tends to a constant. Right: Temporal decay
of $\rho_A$ and $\rho_B$ multiplied by $t^{0.21}$.}
\end{figure}
\begin{figure}
\epsfxsize=130mm \centerline{\epsffile{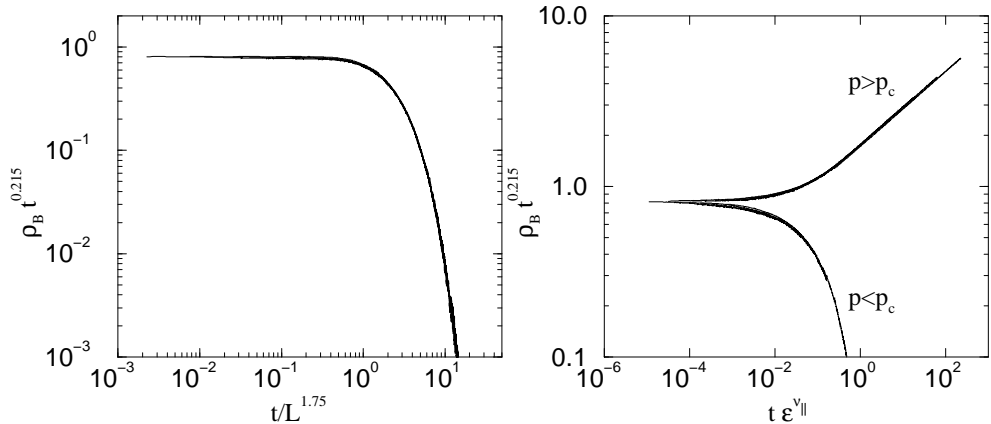}} \vspace{2mm}
\caption{ \label{FIGCOLL} Left: Finite size data collapse
according to the scaling form (\ref{FsForm}). Right: Data collapse
for off-critical simulations according to the scaling
form~(\ref{OffForm}).}
\end{figure}

In order to estimate the critical exponents in 1+1 dimensions, we
perform Monte Carlo simulations. Starting with randomly
distributed $A$'s and $B$'s, we first measure the densities
$\rho_A(t)$ and $\rho_B(t)$ up to $3 \cdot 10^5$ time steps. 
As can be seen in the left panel of Fig.~\ref{FIGDECAY}, the ratio
$\rho_B(t)/\rho_A(t)$ approaches a constant value.
Assuming an algebraic decay, both quantities should therefore 
scale with the same critical exponent
\begin{equation}
\label{delta}
\rho_A(t) \sim  \rho_B(t) \sim t^{-\delta}\,.
\end{equation}
The temporal decay of the particle densities is shown in the right
panel of Fig.~\ref{FIGDECAY}. As in the case of the AF model, we
observe strong corrections to scaling, leading to a considerable
curvature of the data. However, compared to the AF model these
corrections are less severe. Moreover, the curvatures for $\rho_A$
and $\rho_B$ have opposite signs. Thus, the local slopes approach
the postulated `true' value of $\delta$ from both sides. Seeking
for the best compromise, we are able to estimate the critical
points by
\begin{center}
\begin{tabular}{|l||l|l|l|l|}
\hline
$\tau$ & $0$ & $0.1$ & $0.5$ & $1$ \\ \hline
$p_c$ & 0.7674(3) & $0.757(2)$ & $0.6920(1)$ & $0.540(1)$ \\
\hline
\end{tabular}
\end{center}
For $\tau=0.5$ the corresponding exponent is given by
$\delta=\beta/\nu_\parallel=0.21(2)$.
Similar measurements for $\tau=0.1$ and $\tau=1$ (not shown here)
suggest that this value is the same for all $\tau>0$.
In order to obtain the dynamic exponent $z=\nu_\parallel/\nu_\perp$,
we perform finite size simulations at criticality. Here the
particle densities should obey the scaling form
\begin{equation}
\label{FsForm}
\rho_A(t) \sim \rho_B(t)
\sim t^{-\delta} \, f(t/L^z) \,,
\end{equation}
where $f$ is a universal scaling function.
Thus, plotting $\rho t^\delta$ against $t/L^z$,
all data sets should collapse onto a single curve.
Comparing different data collapses it turns out
that $\rho_B$ shows a much cleaner scaling
behavior than $\rho_A$. As shown in Fig.~\ref{FIGCOLL},
the best collapse is obtained for $\delta=0.215(15)$ and
$z=1.75(5)$.

To obtain the third exponent, we study the behavior $\rho_B(t)$
below and above criticality. According to the usual
scaling theory for absorbing-state transition, we expect the scaling
form
\begin{equation}
\label{OffForm}
\rho_B(t) \sim t^{-\delta} \, g(t\,\epsilon^{\nu_\parallel}) \,,
\end{equation}
where $\epsilon=|p-p_c|$ denotes the distance from criticality.
Plotting $\rho_B(t) t^\delta$ against $t\,\epsilon^{\nu_\parallel}$\
(see Fig.~\ref{FIGCOLL}), the best data collapse is obtained for 
$\delta=0.215(20)$ and $\nu_\parallel=1.8(1)$.

\begin{figure}
\epsfxsize=130mm \centerline{\epsffile{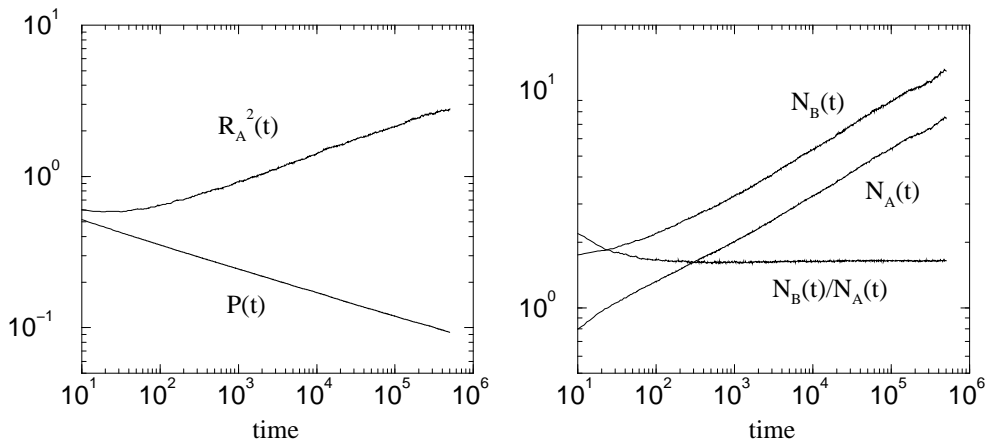}} \vspace{2mm}
\caption{ \label{FIGSEED} Simulations starting with a single
$A$-particle at the origin. Left: Survival probability $P(t)$
and mean square spreading from the origin $R^2(t)$ averaged
over the surviving runs. Right: Average number of particles
$N_A(t)$ and $N_B(t)$ and their quotient.}
\end{figure}

In order to cross-check these estimates,
we perform dynamic simulations starting
with a single pair of particles located
in the center~\cite{GrassbergerTorre79}. As usual in this
type of simulations, we measure the survival probability $P(t)$
that the system has not yet reached one of the two absorbing states,
the average numbers of particles $N_A(t)$ and $N_B(t)$,
and the mean square spreading of all particles
from the origin $R^2(t)$ averaged over the surviving runs.
Assuming that $N_A(t)$ and $N_B(t)$ scale asymptotically
with the same exponent, these quantities should obey
the power laws
\begin{equation}
P(t) \sim t^{-\delta'}\,, \quad
N_A(t) \sim N_B(t) \sim t^{\eta} \,, \quad
R^2(t) \sim t^{2/z}
\end{equation}
with certain dynamical exponents $\delta'$ und $\eta$. As shown in
Fig.~\ref{FIGSEED}, the survival probability shows a clean power
law over almost five decades. Moreover, the quotient
$N_B(t)/N_A(t)$ quickly tends to a constant value. Fitting power
laws to the data shown in Fig.~\ref{FIGSEED}, we obtain the
estimates
\begin{equation}
\delta'=0.15(1)\,, \quad \eta=0.21(1)\,, \quad 2/z=1.16(4)\,.
\end{equation}
Together with the previous results these
estimates satisfy the generalized hyperscaling
relation~\cite{MDHM94}
\begin{equation}
\label{Hyperscaling}
\delta+\delta'+\eta-d/z=0\,
\end{equation}
within numerical errors. Combining all results,
the critical exponents are given by
\begin{equation}
\begin{split}
\beta=0.38(6)\,, & \qquad \beta'=0.27(3) \,, \\
\nu_\parallel=1.8(1) \,, & \qquad \nu_\perp=1.0(1).
\end{split}
\end{equation}
It should be noted that the error bars were obtained by assuming
power-law behavior. Thus, they do not include systematic errors
due to possible corrections to scaling emerging after very long
time.
%
%
%
\section{Conclusions}
\label{Conclusions}
%
%
This work was motivated by recent studies of the annihilation-fission
process which exhibits a novel type of nonequilibrium critical 
behavior. Using a scale-invariant space-time plot 
we have demonstrated that critical clusters 
of the AF process are characterized by an interplay of 
two different dynamic modes. In the high-density mode
we observe spreading avalanches whereas the low-density mode
is characterized by random walks of solitary particles.
In order to understand the interplay between these modes
from a phenomenological point of view,
we have introduced an effective model which involves
two species of particles. The dynamic rules of this model
can be regarded as being composed of cyclically coupled DP 
and pair annihilation processes, i.e., it follows the reaction-diffusion 
scheme $A \leftrightarrow 2A$, $A \rightarrow B$, $2B \rightarrow A$.
The model exhibits a nonequilibrium phase transition which
is in many respects similar to the one observed in the AF process.
In fact, the AF process and the three-state model proposed in the
present work have several features in common:
\begin{itemize}
\item[-] They both have two non-symmetric absorbing states, namely 
the empty lattice and the state with a single diffusing particle.

\item[-] There is no unconventional symmetry (such as parity conservation).

\item[-] Both models exhibit a continuous nonequilibrium phase transition
         with non-DP critical exponents.

\item[-] The visual appearance of critical clusters is very similar
(see Fig.~\ref{FIGDEMO}).
\end{itemize}
Compared to the AF process, the three-state model has several
advantages. On the one hand, it is defined as a
two-site nearest neighbor process. On the other hand, 
corrections to scaling are less severe, allowing
us to determine the critical exponents more accurately.

There are several open questions. Is the critical behavior of the
three-state model universal? If this is indeed the case, does it
represent a independent universality class different from DP and
DP2? Are the AF process and the present model in the same
universality class, i.e., do they have the same critical
exponents? From the field-theoretic point of view, there is no
reason for them to coincide. Yet the two classes may `intersect' 
in 1+1 dimensions. 
Very recently, \'Odor carried out a systematic study of the
annihilation/fission process performing simulations on a parallel computer
combined with generalized mean field approximations and 
coherent anomaly extrapolations~\cite{Odor00}. 
He reports two different universality classes for low and high 
values of the diffusion constant $d$. The corresponding critical 
exponents are shown in Table 1. According to \'Odor, the
exponents for $d \geq 0.5$ are in fair agreement with the exponents
observed in the three-state model. However, this conclusion
is still speculative and needs to be substantiated by further
investigations.

\noindent {\bf Acknowledgements:} I would like to thank
E. Carlon, P. Grassberger, M. Henkel, M.
Howard, J. F. Mendes, G. \'Odor, U. Schollw\"ock, and U. T\"auber
for fruitful discussions.

\newpage

\end{document}